\documentclass[12pt]{article}

\usepackage{amssymb}
\usepackage{amsmath}
\usepackage{amscd}
\usepackage{latexsym}
\usepackage{graphicx}

\usepackage{cite}

\newcommand{\be}{\begin{equation}}
\newcommand{\ee}{\end{equation}}
\newcommand{\Dlt}{\Delta}
\newcommand{\dlt}{\delta}
\newcommand{\prt}{\partial}
\newcommand{\br}{{\bf r}}

\newcommand{\bt}{\beta}
\newcommand{\vp}{\varphi}

\newcommand{\al}{\alpha}
\newcommand{\ra}{\rightarrow}
\newcommand{\sgm}{\sigma}

\newcommand{\gm}{\gamma}
\newcommand{\om}{\omega}
\newcommand{\Om}{\Omega}

\newcommand{\dgr}{\dagger}
\newcommand{\lbd}{\lambda}
\newcommand{\Lbd}{\Lambda}
\newcommand{\rgl}{\rangle}
\newcommand{\lgl}{\langle}

\begin{document}

\begin{center}
 
{\Large{\bf Dynamical Transitions in Trapped Superfluids Excited by Alternating Fields 
} \\ [5mm]

V.I. Yukalov$^{1,2}$ and E.P. Yukalova$^{3}$}  \\ [3mm]

{\it
$^1$Bogolubov Laboratory of Theoretical Physics, \\
Joint Institute for Nuclear Research, Dubna 141980, Russia \\ [2mm]

$^2$Instituto de Fisica de S\~ao Carlos, Universidade de S\~ao Paulo, \\
CP 369, S\~ao Carlos 13560-970, S\~ao Paulo, Brazil \\ [2mm]

$^3$Laboratory of Information Technologies, \\
Joint Institute for Nuclear Research, Dubna 141980, Russia } \\ [3mm]

{\bf E-mails}: {\it yukalov@theor.jinr.ru}, ~~ {\it yukalova@theor.jinr.ru} \\

\end{center}

\vskip 1cm

\begin{abstract}
The paper presents a survey of some dynamical transitions in nonequilibrium trapped 
Bose-condensed systems subject to the action of alternating fields. Nonequilibrium 
states of trapped systems can be realized in two ways, resonant and nonresonant. 
Under resonant excitation, several coherent modes are generated by external alternating 
fields, whose frequencies are tuned to resonance with some transition frequencies of 
the trapped system. A Bose system of trapped atoms with Bose-Einstein condensate can 
display two types of the Josephson effect, the standard one, when the system is separated 
into two or more parts in different locations or when there are no any separation barriers, 
but the system becomes nonuniform due to the coexistence of several coherent modes 
interacting with each other, which is termed internal Josephson effect. The mathematics 
in both these cases is similar. We concentrate on the internal Josephson effect. Systems
with nonlinear coherent modes demonstrate rich dynamics, including Rabi oscillations, 
Josephson effect, and chaotic motion. Under Josephson effect, there exist dynamic transitions 
that are similar to phase transitions in equilibrium systems. The bosonic Josephson effect 
is shown to be realizable not only for weakly interacting systems, but also in superfluids, 
with not necessarily weak interactions. Sufficiently strong nonresonant excitation can 
generate several types of nonequilibrium states comprising vortex germs, vortex rings, 
vortex lines, vortex turbulence, droplet turbulence, and wave turbulence. Nonequilibrium 
states can be characterized and distinguished by effective temperature, effective Fresnel 
number, and dynamic scaling laws. 
\end{abstract}

\vskip 2mm


\section{Introduction}

In recent years, there has been a high interest to the study of dilute gases exhibiting, 
at low temperatures, Bose-Einstein condensation in traps \cite{Parkins_1,Dalfovo_2,
Courteille_3,Andersen_4,Yukalov_5,Bongs_6,Yukalov_7,Posazhennikova_8,Yukalov_9,
Proukakis_10,Yurovsky_11,Yukalov_12,Yukalov_13,Yukalov_14,Yukalov_15} and in optical 
lattices \cite{Moseley_16,Yukalov_17,Krutitsky_18,Yukalov_19}. See also the most recent 
books \cite{Dupuis_2023,Castin_2025}.

In the present paper, we concentrate on nonequilibrium Bose-Einstein condensates created 
by applying external alternating fields. The latter can be of two kinds, resonant and 
nonresonant. The resonant generation of nonequilibrium condensates excites the system 
by means of oscillating fields with the frequencies tuned in resonance with the transition 
frequencies corresponding to the chosen coherent modes. In the nonresonant excitation, 
nonequilibrium condensates are produced by using sufficiently strong alternating fields 
whose frequencies do not need to satisfy some resonant conditions. 
    
When an external potential separates a Bose-condensed system into several clouds in 
different spatial locations, as in optical lattices or in a double well, there can 
arise bosonic Josephson effect, similar to that arising in fermionic Josephson junctions 
\cite{Josephson_20,Josephson_21,Guinea_22,Pashin_23}. When there exists a spatial 
separation of bosonic clouds, as in a double well or in a lattice, one has the usual 
bosonic Josephson effect observed in trapped Bose gases \cite{Cataliotti_24,Albiez_25,Levy_26} 
and in weakly linked reservoirs of superfluid Helium \cite{Pereverzev_27,Sukhatme_28}.   
A setup, realizing an effective double-well potential, can also be created in a binary 
mixture, where two vortices of one component form two effective wells and the other component 
exhibits Josephson oscillations between the cores of the vortices \cite{Bellettini_2024}.  

The other setup is when there exist two interpenetrating populations, not separated 
by any barrier. Then there can appear a current due to the different spatial shapes of 
the populations. The latter effect is called the internal Josephson effect \cite{Ohberg_29} 
or quantum dynamical tunneling \cite{Davis_30}. The mathematics for the both types of 
the Josephson effect is similar. 

In the present paper, we consider the internal Josephson effect in trapped Bose 
systems. The coexistence of several different modes can be realized in the following way. 
In equilibrium systems, Bose-Einstein condensation implies that the lowest energy level 
is occupied by a macroscopic number of atoms. By applying an external resonant field, 
whose frequency $\omega$ is in resonance with the transition frequency $\omega_{12}$ 
between two energy levels, it is possible to create a non-ground-state Bose condensate, 
as is proposed in Refs. \cite{Yukalov_31,Yukalov_32}. To support the existence of 
non-ground-state condensates, one needs to permanently pump energy into the system, 
which becomes nonuniform due to the coexisting condensate modes of different shapes. 
The other possibility is to consider the coexistence of Bose condensates with different 
internal hyperfine states connected with each other by a resonant Rabi field, as is done 
in some experiments \cite{Williams_33,Zibold_34}.  

We consider the internal Josephson effect supported by the resonant generation of nonlinear 
coherent modes in a trapped Bose-condensed system. The modes are termed nonlinear, since 
they represent stationary states of the nonlinear Schr\"{o}dinger equation characterizing 
a Bose-Einstein condensate confined in a trap. And the modes are called coherent as far as 
the condensate wave function describes the coherent part of a Bose-condensed system. 
General information on coherent states can be found in the books \cite{Klauder_35,Mandel_36} 
and in the exhaustive review by Dodonov \cite{Dodonov_37}. Applications of coherent states 
in statistical and Bose-condensed systems are described in Refs. \cite{Yukalov_13,Yukalov_38}.

The nonresonant excitation of a trapped Bose-Einstein condensate creates a nonequilibrium 
system comprising, depending on the strength of the alternating field, different topological
objects, such as vortex germs, vortex rings, and vortex lines, and it can generate highly
nonequilibrium states, as vortex turbulence, droplet turbulence, and wave turbulence.  

The layout of the paper is as follows. In Sec. 2, we briefly recall the derivation of the 
coupled system of equations, necessary for the realization of the bosonic Josephson effect
under weak interactions and zero temperature. Then, in Sec. 3, we describe the dynamical 
transitions occurring in the system and in Sec. 4, we generalize the approach to the case 
of trapped atoms with strong interactions. This generalization demonstrates the possibility 
of observing and studying bosonic Josephson effect, as well as the related dynamical 
transitions, in a larger class of trapped superfluid systems. Section 5 describes some 
ramifications of the resonant method of the coherent modes generation. In particular, the
excitation of coherent modes by means of interaction modulation is discussed. The possibility
of generation of several modes by applying several external oscillating fields is shown. And
the feasibility of higher-order resonances, accompanied by harmonic generation and parametric 
conversion, is referenced. Section 6 is devoted to nonresonant generation of qualitatively
different nonequilibrium states of trapped atoms comprising such topological coherent modes
as vortex germs, vortex rings, vortex lines, and coherent droplets. Different regimes of 
quantum vortex turbulence can be distinguished by the types of dynamic scaling.          

Throughout the paper, the system of units is employed where the Planck, $\hbar$, and 
Boltzmann, $k_B$, constants are set to one.

\section{Resonant Generation}

Let us consider a system of $N$ trapped atoms interacting through the local potential
\be
\label{1}
\Phi(\br) \; = \; \Phi_0 \; \dlt(\br) \qquad 
\left( \Phi_0 = 4\pi\; \frac{a_s}{m}\right) \;   ,
\ee
where $a_s$ is a scattering length and $m$ is atom mass. The second-quantized energy 
Hamiltonian, with the field operators $\psi$, reads as
$$
\hat H \; = \; \int \psi^\dgr(\br,t) \; \left[\; - \; 
\frac{\nabla^2}{2m} + U(\br,t) \; \right] \; \psi(\br,t) \; d\br \; +
$$
\be
\label{2}
+\; \frac{1}{2} \; \Phi_0 \int \psi^\dgr(\br,t) \; \psi^\dgr(\br,t) \; \psi(\br,t) \; 
\psi(\br,t) \; d\br \; ,
\ee
in which the external potential
\be
\label{3}
U(\br,t) \; = \; U(\br) + V(\br,t)
\ee
consists of a trapping potential $U({\bf r})$ and a modulating potential taken in the 
form
\be
\label{4}
 V(\br,t) \; = \; V_1(\br) \cos(\om t) + V_2(\br) \sin(\om,t) \;  .
\ee

At zero temperature and asymptotically weak interactions, all atoms can be assumed 
to be in the condensed state that is a coherent state. The latter is defined as the 
eigenstate of the destruction operator, according to the equation
\be
\label{5}
 \psi(\br,t) \; | \; \eta \; \rgl \; = \; \eta(\br,t) \; | \; \eta \; \rgl \; ,
\ee
where $|\eta>$ is a coherent state, and $\eta_({\bf r},t)$ is a coherent field 
representing the condensate wave function. Averaging the Heisenberg equation of motion 
over the coherent state results in the nonlinear Schr\"{o}dinger equation for the 
condensate wave function
\be
\label{6}
 i\; \frac{\prt}{\prt t} \; \eta(\br,t) \; = \; 
\hat H[\; \eta\; ] \; \eta(\br,t) \;  ,
\ee
with the nonlinear Hamiltonian
\be
\label{7}
 \hat H[\; \eta\; ] \; = \; - \; 
\frac{\nabla^2}{2m} + U(\br,t) + \Phi_0\; | \; \eta(\br,t) \; |^2 \;  .
\ee
Equation (\ref{6}) was advanced by Bogolubov \cite{Bogolubov_39}(see also 
\cite{Bogolubov_40,Bogolubov_41}) and later studied in many works starting with 
Gross \cite{Gross_42,Gross_43,Gross_44,Gross_45,Gross_46}, Pitaevskii \cite{Pitaevskii_47},
and Wu \cite{Wu_48}.  

The time-dependent wave function $\eta({\bf r},t)$ can be expanded 
\cite{Yukalov_49,Yukalov_50} over the stationary coherent modes $\varphi_n({\bf r})$ 
satisfying the equation
\be
\label{8}
 H[\; \vp_n \; ] \; \vp_n(\br) \; = \; E_n \vp_n(\br) \; ,
\ee
where
$$
H[\; \vp_n \; ] \; = \; - 
\frac{\nabla^2}{2m} + U(\br) + N \Phi_0\; | \; \vp_n(\br) \; |^2 \;  .
$$
This expansion reads as
\be
\label{9}
 \eta(\br,t) \; = \; \sqrt{N} \; \sum_n c_n(t) \; e^{-i E_n t} \; \vp_n(\br) \;  .
\ee
The coherent modes are not necessarily mutually orthogonal, but they are normalized, 
\be
\label{10}
\int \vp_n^*(\br) \; \vp_n(\br) \; d\br \; = \; 1 \;   .
\ee

Suppose, we wish to generate two coherent modes with energies $E_1 < E_2$, with the 
transition frequency
\be
\label{11}
\om_{21} \; \equiv \; E_2 - E_1 \;   .
\ee
For this purpose, the frequency of the modulating field (\ref{4}) has to be close to 
the transition frequency $\omega_{21}$, such that the quasi-resonance condition
\be
\label{12}
 \left| \; \frac{\Dlt\om}{\om} \; \right| \; \ll \; 1 \qquad
(\Dlt\om \equiv \om - \om_{21} )  
\ee
be valid.

Transitions between the energy levels are induced by two transition amplitudes, one 
being caused by atomic interactions, 
\be
\label{13}
\al_{mn} \; \equiv \; \Phi_0 N \int |\; \vp_m(\br) \; |^2  \left\{ 
2 |\; \vp_n(\br) \; |^2  - |\; \vp_m(\br) \; |^2 \right\} \; d\br \; ,
\ee
and the other being due to the alternating field with the Rabi frequency
\be
\label{14}
\bt_{mn} \; \equiv \; 
\int  \vp_m^*(\br) \; [\; V_1(\br) - i V_2(\br) \; ] \; \vp_n(\br) \; d\br \;  .
\ee
In order to preserve good resonance and to avoid power broadening, the transition 
amplitudes need to be smaller than the transition frequency,
\be
\label{15}
 \left| \; \frac{\al_{12}}{\om_{12}} \; \right| \; \ll \; 1 \; , 
\qquad
\left| \; \frac{\al_{21}}{\om_{21}} \; \right| \; \ll \; 1 \; , 
\qquad 
\left| \; \frac{\bt_{12}}{\om_{12}} \; \right| \; \ll \; 1 \; .
\ee

If at the initial moment of time only two modes are populated, then switching on 
the resonant field does not populate other energy levels, leaving touched only the 
chosen two modes, which results in the equations for the coefficients $c_n(t)$ of 
expansion (\ref{9}) in the form
$$
i \; \frac{dc_1}{dt} \; = \; 
\al_{12} \; |\; c_2 \; |^2 c_1 + 
\frac{1}{2}\; \bt_{12}\; c_2 \; e^{i\Dlt\om \cdot t} \; ,
$$
\be
\label{16}
 i \; \frac{dc_2}{dt} \; = \; 
\al_{21} \; |\; c_1 \; |^2 c_2 + 
\frac{1}{2}\; \bt_{12}^*\; c_1 \; e^{-i\Dlt\om \cdot t} \;  ,
\ee
with the normalization constraint
\be
\label{17}
|\; c_1 \; |^2 + |\; c_2 \; |^2 \; = \; 1 \;  .
\ee    

It is convenient to introduce the population difference
\be
\label{18}
s \; \equiv \; |\; c_2 \; |^2 - |\; c_1 \; |^2 \;   ,
\ee
using which the coefficient functions $c_i$ can be represented as
$$
c_1 \; = \; \sqrt{ \frac{1-s}{2} } \; 
\exp\left\{ i \; \left( \pi_1 + \frac{\Dlt\om}{2} \; t \right) \right\} \; ,
$$
\be
\label{19}
c_2 \; = \; \sqrt{ \frac{1+s}{2} } \; 
\exp\left\{ i \; \left( \pi_2 - \frac{\Dlt\om}{2} \; t \right) \right\} \; ,
\ee
where $\pi_i = \pi_i(t)$ are real-valued phases.

Introduce also the average interaction amplitude
\be
\label{20}
\al \; \equiv \; \frac{1}{2} \; ( \al_{12} + \al_{21} ) \;   ,
\ee
the notation for the Rabi frequency
\be
\label{21}
\bt_{12} \; \equiv \; \bt \; e^{i\nu} \qquad 
( \bt \equiv |\;\bt_{12}\;| ) \;   ,
\ee
and the effective detuning 
\be
\label{22}
\dlt \; \equiv \; \Dlt\om + \frac{1}{2} \; ( \al_{12} - \al_{21} ) \; .
\ee

The effective phase difference is denoted as
\be
\label{23}
x \; \equiv \; \pi_2 - \pi_1 + \nu \;  .
\ee
The population difference $s$ and the phase difference $x$ are the main quantities defining
the dynamics of the generated populations. These functions of time, varying in the intervals
\be
\label{24}
s \; \in \; [-1, \; 1 ] \; , \qquad  x \; \in \; [0,2\pi) \; ,
\ee
satisfy the equations
\be
\label{25}
\frac{ds}{dt} \; = \; - \bt\; \sqrt{1-s^2} \; \sin x \; , \qquad 
\frac{dx}{dt} \; = \; \al s + \frac{ \bt s}{ \sqrt{1-s^2} } \; \cos x \; + \dlt \;   .
\ee
The latter can be obtained from the Hamiltonian equations
\be
\label{26}
 \frac{ds}{dt} \; = \; - \; \frac{\prt \overline H}{\prt x} \; ,
\qquad
\frac{dx}{dt} \; = \; \frac{\prt \overline H}{\prt s} \;  ,
\ee
with the effective Hamiltonian
\be
\label{27}
\overline H \; = \; \frac{1}{2} \; \al s^2 - \bt \; \sqrt{1 - s^2} \; \cos x +
\dlt s \;   .
\ee

Employing the dimensionless quantities for the Rabi frequency and effective detuning,
\be
\label{28}
b \; \equiv  \; \frac{\bt}{\al} \; , \qquad \epsilon \; = \; \frac{\dlt}{\al} \;  ,
\ee
and measuring time in units of $1/\alpha$, reduces the effective Hamiltonian to 
\be
\label{29}
 H(s,x) \; \equiv \; \frac{\overline H}{\al} \; = \; 
 \frac{1}{2} \; s^2 - b \; \sqrt{1 - s^2} \; \cos x +
\epsilon s \;  ,
\ee
and the dynamical equations to the form
\be
\label{30}
\frac{ds}{dt} \; = \; - b \; \sqrt{1-s^2} \; \sin x \; , \qquad 
\frac{dx}{dt} \; = \;  s + \frac{ b s}{ \sqrt{1-s^2} } \; \cos x \; + \epsilon \;   .
\ee

\section{Dynamical Transitions}

Under a dynamical transition one understands the qualitative change of the phase portrait 
formed by the set of fixed points \cite{Nemytskii_51,Temam_52}. Dynamic and stationary 
solutions of the evolution equations, similar to (\ref{30}), have been considered in several 
papers \cite{Yukalov_31,Yukalov_32,Yukalov_49,Yukalov_50,Marzlin_53,Marzlin_54,Yukalov_55,
Ostrovskaya_56,Williams_57,Yukalov_58,Kivshar_59,Agosta_60,Yukalov_61,Yukalov_62} and
summarized in the reviews \cite{Yukalov_12,Yukalov_14}. Here we follow the analysis of the 
papers \cite{Yukalov_31,Yukalov_32}. We keep in mind a small detuning, such that 
$|\epsilon| \ll 1$.      

Depending on the parameter $b$, characterizing the dimensionless Rabi frequency normalized 
by the interaction strength ($b = \beta/\alpha$), there can exist several dynamic regimes. 
When the Rabi frequency is larger than the interaction strength, there are two fixed points:
$$
s_1^* \; = \; \frac{\epsilon}{b} \; , \qquad x_1^* \; = \; 0 \; ,
$$
\be
\label{31}
 s_2^* \; = \; -\; \frac{\epsilon}{b} \; , \qquad x_2^* \; = \; \pi
\qquad ( b^2 \geq 1)  \;  ,
\ee
both being the centers. This regime is close to the regime of classical Rabi oscillations.

When the Rabi frequency is smaller than the interaction strength, with the interactions being 
repulsive, then there are four fixed points:
$$
s_1^* \; = \; \frac{\epsilon}{b} \; , \qquad x_1^* \; = \; 0 \; ,
$$
$$
s_2^* \; = \; -\; \frac{\epsilon}{b} \; , \qquad x_2^* \; = \; \pi \; ,
$$
$$
s_3^* \; = \; \sqrt{1-b^2} + \frac{b^2\epsilon}{1-b^2} \; , \qquad x_3^* \; = \; \pi \; ,
$$
\be
\label{32}
 s_4^* \; = \; - \sqrt{1-b^2} + \frac{b^2\epsilon}{1-b^2} \; , \qquad x_4^* \; = \; \pi 
\qquad ( 0 < b < 1 ) \;  .
\ee
The fixed point $\{s^*_2,x^*_2\}$ is a saddle, while all other points are centers. 

If the interactions are attractive, then the parameter $b < 0$ is negative, and instead of the 
previous fixed points, one has:
$$
s_1^* \; = \; \frac{\epsilon}{b} \; , \qquad x_1^* \; = \; 0 \; ,
$$
$$
s_2^* \; = \; -\; \frac{\epsilon}{b} \; , \qquad x_2^* \; = \; \pi \; ,
$$
$$
s_5^* \; = \; \sqrt{1-b^2} + \frac{b^2\epsilon}{1-b^2} \; , \qquad x_5^* \; = \; 0 \; ,
$$
\be
\label{33}
 s_6^* \; = \; - \sqrt{1-b^2} + \frac{b^2\epsilon}{1-b^2} \; , \qquad x_6^* \; = \; 0
\qquad ( -1 < b < 0 ) \;  .
\ee
As is seen, the dynamics is similar to the previous case, except that the phase difference is 
shifted by $\pi$. This happens due to the symmetry of the evolution equations (\ref{30}) with
respect to the replacement $b \mapsto -b$, $x \mapsto x - \pi$, and 
$\epsilon \mapsto - \epsilon$. Therefore, it is admissible to limit the consideration by 
the positive $b > 0$. Let us recall that $b = 0$ means the absence of pumping and no mode 
generation, with $s$ being constant.   
 
In the case of a pure resonance, when $\omega = \omega_{21}$, hence $\epsilon = 0$, the 
dynamical transition, happening at $b = 1$, is the transition from the set of two stable 
centers
$$
s_1^* \; = \; 0 \; , \qquad x_1^* \; = \; 0 \qquad (center) \; ,
$$
\be
\label{34}
s_2^* \; = \; 0 \; , \qquad x_2^* \; = \; \pi \qquad (center)  \qquad ( b^2 \geq 1) \; ,
\ee
to the set of three centers and one saddle point
$$
s_1^* \; = \; 0 \; , \qquad x_1^* \; = \; 0 \qquad (center) \; ,
$$
$$
s_2^* \; = \; 0 \; , \qquad x_2^* \; = \; \pi \qquad (saddle) \; ,
$$
$$
s_3^* \; = \; \sqrt{1 - b^2}  \; , \qquad x_3^* \; = \; \pi \qquad (center) \; ,
$$
\be
\label{35}
 s_4^* \; = \; - \sqrt{1 - b^2}  \; , \qquad x_4^* \; = \; \pi \qquad (center) 
\qquad ( 0 < b < 1 ) \;  .
\ee
The regime with $b < 1$, where the Rabi frequency is smaller than the interaction strength,
is sometimes associated with the Josephson dynamics. The change of the dynamic regime at 
$b = 1$ is named a pitchfork bifurcation. The variation of the fixed point $s^*$ as a function
of $b$ is shown in Fig. 1.

\begin{figure}[ht]
\centerline{\includegraphics[width=10cm]{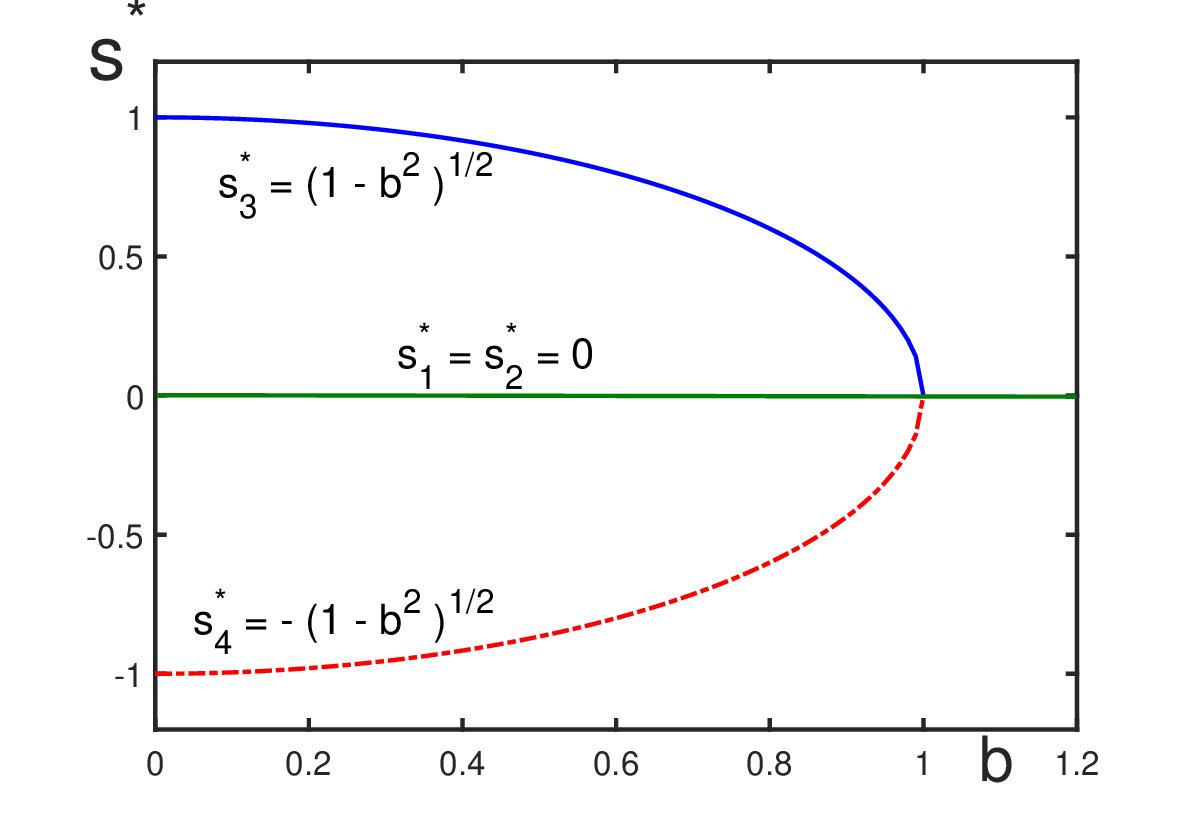}}
\caption{\small 
Pitchfork bifurcation. Fixed point $s^*$ characterizing the mode population difference, 
as a function of the parameter $b$.} 
\label{fig:Fig.1}
\end{figure}

Since $b = \beta/\alpha$ is the ratio of the Rabi frequency $\beta$ to the interaction 
strength $\alpha$, the regime with $b > 1$ can be called the regime of strong pumping or 
weak interactions, while that with $b < 1$ as the regime of weak pumping or strong 
interactions. Sometimes, one names the regime of $b > 1$ as the regime of Rabi dynamics 
and that of $b < 1$, the regime of Josephson dynamics, although for all $b > 0$ the 
approximate solutions for the mode populations can be represented \cite{Yukalov_31} as
\be
\label{36}
 n_1 \; = \; 1 \; - \; \frac{|\;\bt\; |^2}{\Om^2} \; \sin^2 \frac{\Om t}{2} \;  ,
\qquad 
 n_2 \; = \;  \frac{|\;\bt\; |^2}{\Om^2} \; \sin^2 \frac{\Om t}{2} \;
\ee
with the effective Rabi frequency $\Omega$ defined by the expression
\be
\label{37}
 \Om^2 \; = \; [\; \al ( n_1 - n_2) - \Dlt\om \; ]^2 + |\; \bt \; |^2 \;  ,
\ee
where the notation 
\be
\label{38}
n_1 \; \equiv \; |\; c_1 \; |^2 \; , \qquad
n_2 \; \equiv \; |\; c_2 \; |^2 \; ,
\ee
is used and the initial condition
\be
\label{39}
c_1(0) \; = \; 1 \; , \qquad c_2(0) \; = \; 0 \; ,
\ee
is accepted. Thus the dynamics is always of the effective Rabi type, if to keep in mind the 
effective Rabi frequency (\ref{37}). However, the latter, depends on the relation between 
the standard Rabi frequency $\beta$ and the interaction strength $\alpha$ and it is a function 
of time. For the convenience of nomenclature, we may also term the regime of $b > 1$, the Rabi 
regime and that of $b < 1$, the Josephson regime.    

In addition to the dynamical transition between the Rabi and Josephson dynamics, there is one 
more nonstandard dynamical transition related to the effect of saddle separatrix crossing 
\cite{Yukalov_14,Yukalov_31,Yukalov_32}. In the Josephson regime, the fixed point 
$\{s^*_2,x^*_2\} = \{0,\pi\}$ is a saddle. The trajectory traversing the saddle is called 
the saddle separatrix, since it separates the basins of attraction for different fixed points. 
The separatrix satisfies the condition
\be
\label{40}
 H(s,x) \; = \; H(s_2^*,x_2^*) \;  ,
\ee
which yields the separatrix equation
\be
\label{41}
\frac{1}{2} \; s^2 - b \sqrt{1 - s^2} \; \cos x + \epsilon s \; = \; b \;   .
\ee
The separatrix is shown in Fig. 2 for the case of the resonance, when $\epsilon = 0$, and 
in Fig. 3, for the detuning parameter $\epsilon = 0.1$. 

\begin{figure}[ht]
\centerline{\includegraphics[width=10cm]{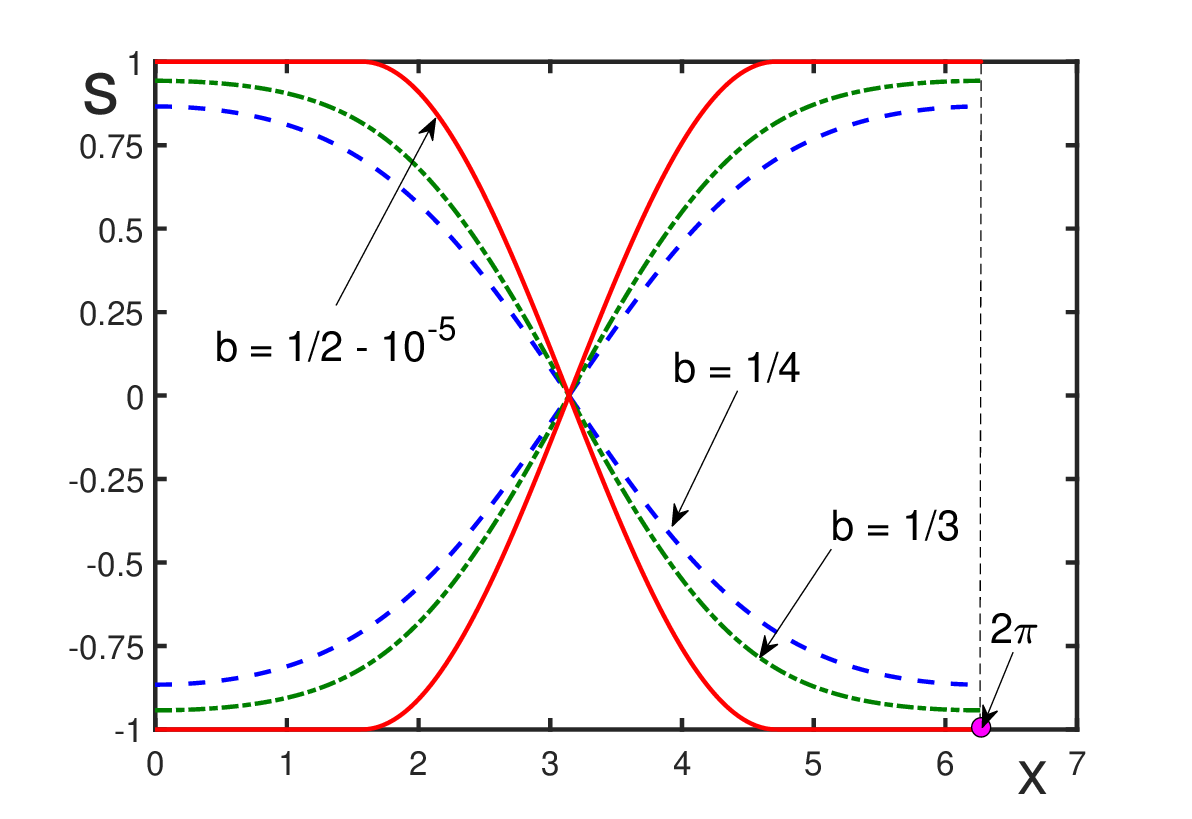}}
\caption{\small 
Behavior of the separatrix branches, positive, $s(x)>0$, and negative, $s(x)<0$, for the case
of pure resonance, where $\epsilon=0$, as functions of the phase difference $x\in[0,2\pi]$ 
for $b=1/4$ (dashed line), $b=1/3$ (dashed-dotted line), and $b=1/2-10^{-5}$ (solid line)} 
\label{fig:Fig.2}
\end{figure}

\begin{figure}[ht]
\centerline{
\hbox{ \includegraphics[width=8cm]{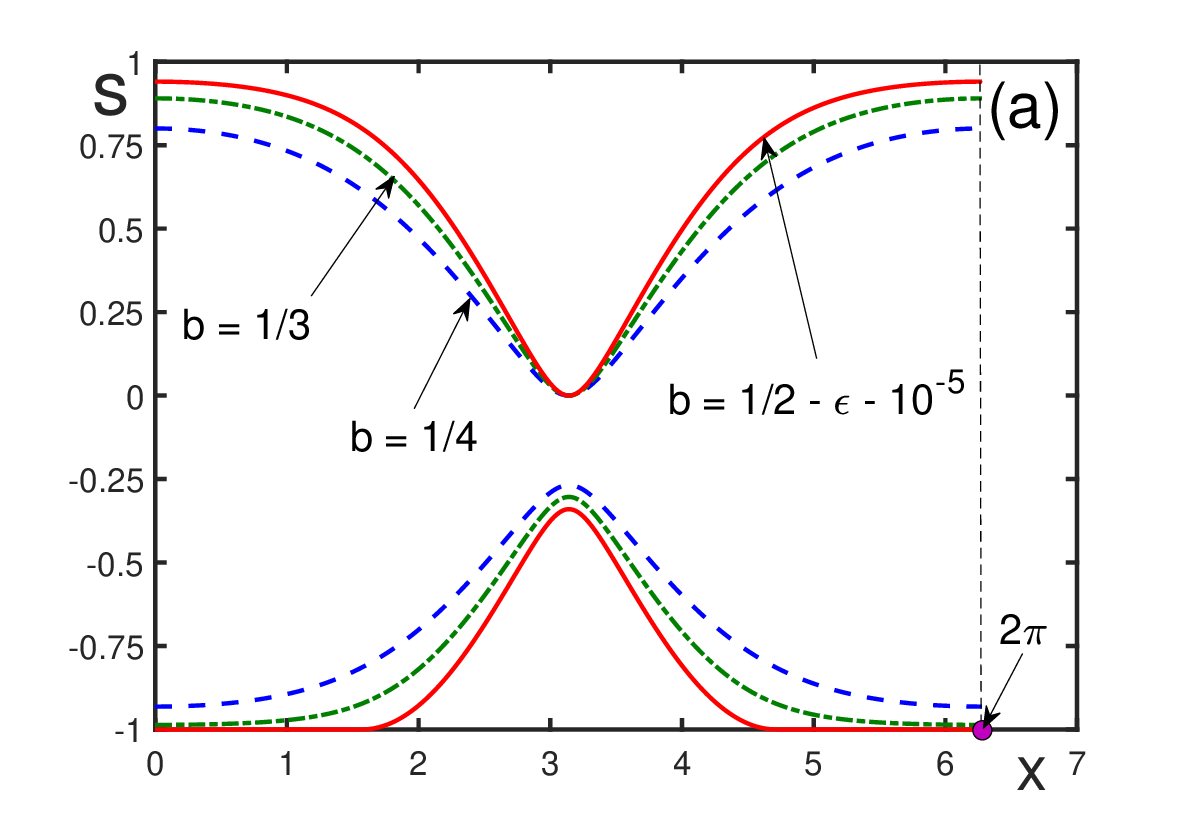}\hspace{1cm}
\includegraphics[width=8cm]{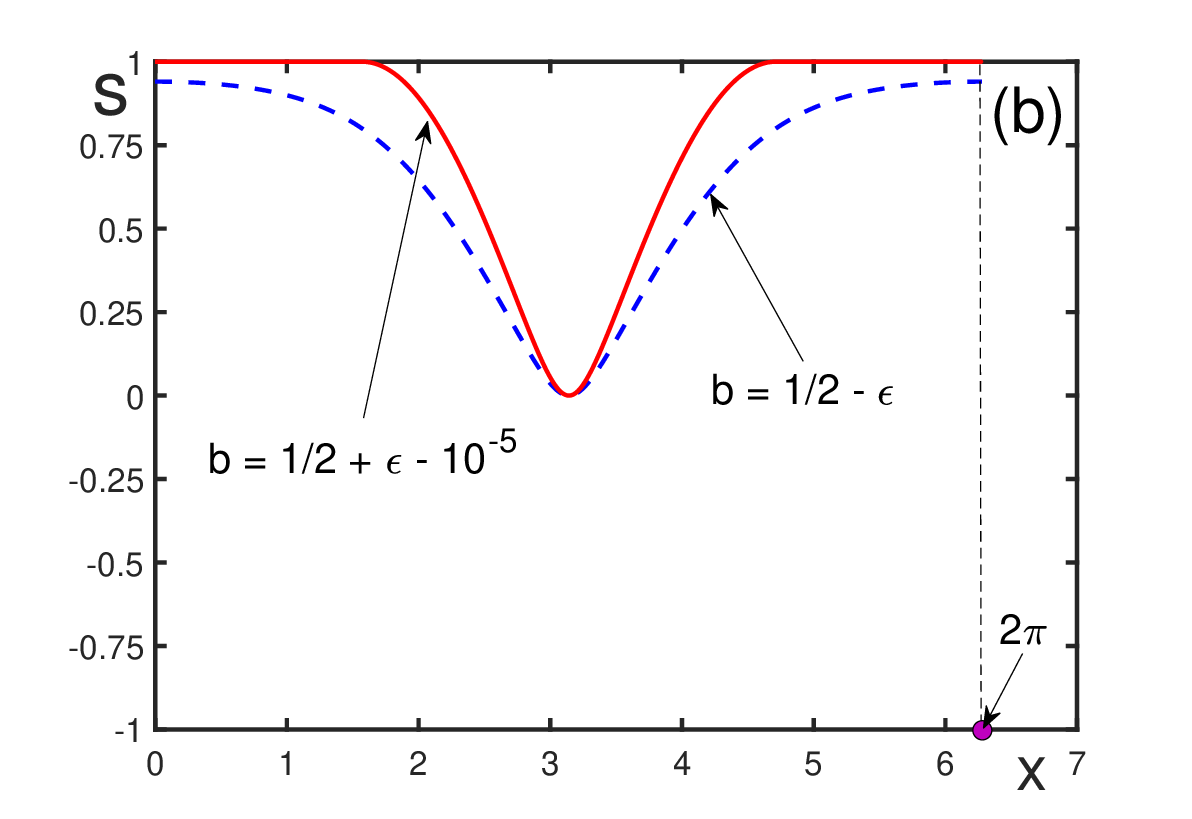}  } }
\caption{\small 
Behavior of the separatrix $s(x)$ for the detuning $\epsilon=0.1$ as a function of the
phase difference $x\in[0,2\pi]$.
(a) The separatrix branches $s(x)>0$ and $s(x)<0$ for $b=1/4$ (dashed line), $b=1/3$ 
(dashed-dotted line), and $b=1/2-\epsilon-10^{-5}$ (solid line); 
(b) The upper separatrix branch $s(x)>0$, for $b=1/2-\epsilon$ (dashed line) and for 
$b=1/2+\epsilon-10^{-5}$ (solid line). For $b>1/2-\epsilon$, the low branch $s(x)<0$  
does not exist.
}
\label{fig:Fig.3}
\end{figure}

Actually, the separatrix consists of two parts, the upper and the lower ones. If the initial 
conditions $s_0$ and $x_0$ are above the upper separatrix, the trajectory is always locked 
from below by the upper separatrix, And if the initial conditions are below the lower 
separatrix, the trajectory is always locked from above by the lower separatrix. This is 
called the effect of mode locking \cite{Yukalov_14,Yukalov_31,Yukalov_32}. The critical line 
on the parametric plane $\{b,\epsilon\}$, where the effect of the transition between the 
mode-locked and mode-unlocked regimes happens, is the separatrix line crossing the initial 
point of the trajectory, so that
\be
\label{42}
 H(s_2^*,x_2^*) \; = \; H(s_0,x_0) \;  ,
\ee
which yields the critical-line equation
\be
\label{43}
\frac{1}{2} \; s_0^2 - b_c \sqrt{1 - s_0^2} \; \cos x_0 + \epsilon_c s_0 \; = \; b_c \; .
\ee
The critical parameter $b_c$ is expressed through $s_0$ and $x_0$ as
\be
\label{44}
 b_c \; = \; \frac{s_0^2 + 2\epsilon_c s_0}{2(1+\sqrt{1-s_0^2}\; \cos x_0) } \;  .
\ee
For $b < b_c$, depending on the chosen initial condition, the mode locking implies that the 
trajectory of the population difference $s$ is locked in the lower or upper half of the 
interval $[-1,1]$, so that, e.g., either 
\be
\label{45}
- 1 \; \leq \; s \; < \; 0 \qquad ( b < b_c, \; s_0 = -1 ) \;   ,
\ee
or
\be
\label{46}
0 \; \leq \; s \; < \; 1 \qquad ( b < b_c, \; s_0 = 1 ) \;    .
\ee
But when $b > b_c$, the population difference oscillates in the whole interval $[-1,1]$,
thus the mode being not locked. 

For example, under the initial condition $s_0 = \pm 1$, the critical line simplifies to
\be
\label{47}
b_c \; = \; \frac{1}{2} \pm \epsilon_c \;  .
\ee
When crossing the critical line, the dynamics change from the mode-locked to mode-unlocked 
regimes and the oscillation period doubles. The critical dynamics in the vicinity of the 
critical line is studied in Refs. \cite{Yukalov_14,Yukalov_31,Yukalov_32}. 

Summarizing, the dynamics of the coherent-mode generation, under the initial condition
with $s_0 = \pm 1$, contains the following regimes:
$$
b \; = \; 0 \qquad ( equilibrium) \; ,
$$
$$
0 \; < \; b \; < \; b_c  \qquad ( mode-locked \; Josephson \; regime) \; ,
$$
$$
b \; = \; b_c \qquad (critical \; dynamics) \; ,
$$
$$
b_c \; < \; b \; < \; 1  \qquad ( mode-unlocked\; Josephson \; regime) \; ,
$$
$$
b \; = \; 1 \qquad (pitchfork \; bifurcation) \; ,
$$
\be
\label{48}
b \; > \; 1 \qquad (Rabi \; regime) \;   .
\ee 
When increasing the pumping parameter $b$, the system passes from the mode-locked regime 
to the mode-unlocked regime. Crossing the critical line, with increasing $b$, the system 
dynamics changes dramatically. In particular, the oscillation amplitude and the oscillation 
period approximately double \cite{Yukalov_14}.

The dynamic transition on the critical line is similar to a phase transition in an 
equilibrium statistical system \cite{Yukalov_31,Yukalov_32,Yukalov_58,Yukalov_63}. To show 
this, we need to define an effective stationary energy. For this purpose, we notice that 
the evolution equations (\ref{16}) can be represented in the Hamiltonian form
\be
\label{49}
 i\; \frac{dc_1}{dt} \; = \; \frac{\prt H_{eff}}{\prt c_1^*} \; ,
\qquad
 i\; \frac{dc_2}{dt} \; = \; \frac{\prt H_{eff}}{\prt c_2^*} \; ,
\ee
with the effective Hamiltonian
\be
\label{50}
H_{eff} \; = \; \al n_1 n_2 + \frac{1}{2} \; \left( \bt \; e^{i\Dlt\om \cdot t} \; c_1^* c_2 +
\bt^* \; e^{-i\Dlt\om \cdot t} \; c_2^* c_1 \right) \;  .
\ee

Equations (\ref{16}), in the Josephson regime, where $b < 1$, can be solved resorting to the
averaging techniques \cite{Bogolubov_64} and the scale separation approach \cite{Yukalov_65}.
Then the functional variables $n_i$ are treated as slow, as compared to the fast variables
$c_i$. Using this and averaging the effective Hamiltonian over time yields the effective 
energy
\be
\label{51}
E_{eff} \; \equiv \; \overline H_{eff} \; = \;
\frac{\al b^2}{2 u^2} \; \left( \frac{b^2}{2u^2} + \dlt \right) \;   ,
\ee
where the bar means time averaging and 
$$
u^2 \; = \; \frac{1}{2} \; \left( 1 + \sqrt {1 - 4b^2} \right) \;  .
$$

The {\it order parameter} can be defined as the time-averaged population difference
\be
\label{52}
 \eta  \; \equiv \; \overline n_1 - \overline n_2 \; = \; 1 \; - \; \frac{b^2}{u^2} \;  .
\ee
The {\it pumping capacity}, describing the capacity of the system to store the energy 
pumped into it, is given by the expression
\be
\label{53}
C_\bt  \; \equiv \; \frac{\prt E_{eff}}{\prt \bt} \;  .
\ee
The dependence of the order parameter on the detuning is characterized by the {\it detuning
susceptibility}
\be  
\label{54}
 \chi_\dlt \; \equiv \; \left| \; \frac{\prt\eta}{\prt\dlt} \; \right| \;  .
\ee 

In the vicinity of the critical pumping $b_c$, given by Eq. (\ref{47}), the above 
characteristics display the critical behavior with respect to the diminishing variable
\be
\label{55}
\tau  \equiv \; | \; b - b_c \; | \;  .
\ee 
This behavior is defined by the asymptotic expressions for the order parameter
\be
\label{56}
 \eta \; \simeq \; \frac{1}{\sqrt{2}} \; ( 1 - 2 \dlt ) \; \tau^{1/2} \qquad 
(\tau\ra 0) \;  ,
\ee
pumping capacity
\be
\label{57}
 C_\bt  \; \simeq \; \frac{1}{4\sqrt{2}} \; \tau^{-1/2} \qquad 
(\tau\ra 0) \;   ,
\ee
and the detuning susceptibility
\be
\label{58}
 \chi_\dlt  \; \simeq \; \frac{1}{\sqrt{2}} \; \tau^{-1/2} \qquad 
(\tau\ra 0) \;   .
\ee
The related critical exponents satisfy the same sum rule as in the case of equilibrium 
statistical systems: $0.5 + 2 \times 0.5 + 0.5 = 2$.

\section{Strong Interactions} 

The dynamics studied above assumes that the system is gaseous, being composed of very 
weakly interacting particles, such that these interactions are so weak, that all particles
in the system are Bose-condensed and the interactions do not disturb much the evolution 
of the condensed part of the system. In order to study the influence of interactions in 
a more general situation, one has, first of all, to address the general form of the equation 
for the condensate. In the present section, we analyze the possibility of realizing the 
Josephson effect in a Bose-condensed system of particles with sufficiently strong 
interactions.  

The occurrence of condensate implies that one has to consider a system with global gauge 
symmetry breaking \cite{Yukalov_5,Yukalov_9,Yukalov_12,Yukalov_13,Lieb_66,Yukalov_67}.   
Generally, to consider a Bose system with broken gauge symmetry, it is sufficient to employ
the Bogolubov shift for the field operator
\be
\label{59}
 \psi(\br,t) \; = \; \eta(\br,t) + \psi_1(\br,t) \;  ,
\ee
where $\eta$ is the condensate wave function and $\psi_1$ is an operator of uncondensed 
particles. These variables are mutually orthogonal,
\be
\label{60}
 \int \eta^*(\br,t) \; \psi(\br,t) \; d\br \; = \; 0 \;  .
\ee

The grand Hamiltonian takes the form
\be
\label{61}   
H \; = \;  \hat H - \mu_0 N_0 - \mu_1 \hat N_1 - \hat\Lbd \; ,
\ee
in which $\hat{H}$ is the energy Hamiltonian (\ref{2}), $\mu_0$ is a condensate chemical 
potential guaranteeing the normalization
\be
\label{62}
N_0 \; = \; \int |\; \eta(\br,t) \; |^2 \; d\br \;   ,
\ee
to the number of condensed particles, $\mu_1$ is the chemical potential of uncondensed 
particles preserving the normalization
\be
\label{63}
 N_1 \; = \; \lgl \; \hat N_1 \; \rgl \; , \qquad
\hat N_1 \; = \; \int \psi_1^\dgr(\br,t) \; \psi_1(\br,t) \; d\br \; ,
\ee
to the number of uncondensed particles, and the operator
\be
\label{64}
\hat\Lbd  \; = \; \int \left[\; \lbd(\br,t) \;\psi_1^\dgr(\br,t) + 
\lbd^*(\br,t) \;\psi_1(\br,t) \; \right] \; d\br
\ee
controls the quantum-number conservation condition
\be
\label{65}
\lgl \; \hat \Lbd \; \rgl \; = \; 0 \;  .
\ee
For the purpose of the latter, the Lagrange multipliers $\lambda$ are chosen so that to 
cancel in the Hamiltonian (\ref{61}) the terms linear in the operators $\psi_1$. 

The condensate wave function is defined by the equation
\be
\label{66}
 i\; \frac{\prt}{\prt t} \; \eta(\br,t) \; = \; \left\lgl \;
\frac{\dlt H}{\dlt \eta^*(\br,t)}\;\right\rgl \;   , 
\ee
which gives the condensate equation
$$
 i\; \frac{\prt}{\prt t} \; \eta(\br,t) \; = \;
\left( - \; \frac{\nabla^2}{2m} + U - \mu_0 \right) \;\eta(\br,t) \; +
$$
\be
\label{67}
 +\;
\int \Phi(\br-\br') \; [\; \rho(\br',t) \;\eta(\br,t) +
\rho_1(\br,\br',t) \;\eta(\br',t) + \sgm_1(\br,\br',t) \; \eta^*(\br',t) +
\xi_1(\br,\br',t) \; ]\; d\br' \; .
\ee

Here and in what follows the notations are used for the particle density 
\be
\label{68}
 \rho(\br,t) \; = \; \rho_0(\br,t) + \rho_1(\br,t) \;  ,
\ee
the density of condensed particles 
\be
\label{69}
 \rho_0(\br,t) \; \equiv \; |\; \eta(\br,t) \; |^2 \;  ,
\ee
the density of uncondensed particles
\be
\label{70}
 \rho_1(\br,t) \; \equiv \; \rho_1(\br,\br,t) \; = \;
\lgl \;\psi_1^\dgr(\br,t) \;\psi_1(\br,t) \; \rgl \;  ,
\ee
the single-particle density matrix
\be
\label{71}
  \rho_1(\br,\br',t) \; = \; \lgl \;\psi_1^\dgr(\br',t) \;\psi_1(\br,t) \; \rgl \;  ,
\ee
the amplitude of pairing particles (so-called anomalous average)
\be
\label{72}
\sgm_1(\br,t) \; \equiv \; \sgm_1(\br,\br,t) \; = \;
\lgl \;\psi_1(\br,t) \;\psi_1(\br,t) \; \rgl \;   ,
\ee
where
\be
\label{73}
 \sgm_1(\br,\br',t) \; = \; \lgl \;\psi_1(\br',t) \;\psi_1(\br,t) \; \rgl \; ,
\ee
and the triple anomalous average 
\be
\label{74}
\xi_1(\br,\br',t) \; = \; \lgl \; \psi_1^\dgr(\br',t) \; \psi_1(\br',t) \;
\psi_1(\br,t) \; \rgl \; .
\ee
 
For the local interaction potential (\ref{1}), the condensate equation (\ref{67}) reads 
as
$$
i \; \frac{\prt}{\prt t} \; \eta(\br,t) \; = \; \left( -\;\frac{\nabla^2}{2m} +
U - \mu_0 \right) \; \eta(\br,t) \; +
$$
\be
\label{75}
+ \; \Phi_0 \left\{\; \left[ \; \rho_0(\br,t) + 2\rho_1(\br,t) \; \right] \;
\eta(\br,t) + \sgm_1(\br,t) \; \eta^*(\br,t) + \xi_1(\br,t)\; \right\} \; ,
\ee
where
\be
\label{76}
 \xi_1(\br,t) \; \equiv \xi_1(\br,\br,t) \;  .
\ee
If we use the Hartree-Fock-Bogolubov approximation, the latter expression becomes zero, 
although generally it is finite.

In equilibrium, the functions entering equation (\ref{75}) do not depend on time. 
Then, introducing the supplementary Hamiltonian $H_{sup}[\eta]$ acting on the condensate 
function according to the definition
$$
H_{sup}[\; \eta \; ] \; \eta(\br) \; = \; \left[ \; - \frac{\nabla^2}{2m} +
U(\br) \; \right] \; \eta(\br) \; +
$$
\be
\label{77}
+ \; \Phi_0 \left\{\; \left[ \; |\; \eta(\br) \; |^2 + 2\rho_1(\br) \; \right]\;
\eta(\br) + \sgm_1(\br) \; \eta^*(\br) + \xi_1(\br)\; \right\} \; ,
\ee
equation (\ref{67}), in the absence of the external perturbation (\ref{4}), reduces to 
the equilibrium eigenvalue form
\be
\label{78}
 H_{sup}[\; \eta \; ] \; \eta(\br) \; = \; \mu_0 \; \eta(\br) \;  .
\ee

Generally, the eigenvalue equation of type (\ref{78}) can lead to a set of stationary 
solutions of the equation
\be
\label{79}
H_{sup}[\; \eta_n \; ] \; \eta_n(\br) \; = \; E_n \; \eta_n(\br) \; ,
\ee
with the lowest energy level corresponding to the chemical potential,
\be
\label{80}
 \mu_0 \; = \; \min_n E_n \;  .
\ee
 
To compare the problem with the zero-temperature case, it is convenient to pass from the 
functions $\eta_n$, normalized to $N_0$, to the functions $\varphi_n$, normalized to one, by
the substitution
\be
\label{81} 
\eta_n(\br) \; = \; \sqrt{N_0} \; \vp_n(\br) \; ,
\ee
in which
\be
\label{82}
 \int |\; \eta_n(\br) \; |^2 \; d\br \; = \; N_0 \; ,
\qquad
  \int |\; \vp_n(\br) \; |^2 \; d\br \; = \; 1 \; .
\ee
By defining the supplementary Hamiltonian $H_{sup}[\varphi_n]$ by its action
$$
H_{sup}[\; \vp_n\; ] \; \vp_n(\br) \; = \; \left[ \; -
\frac{\nabla^2}{2m} + U(\br) \; \right] \; \vp_n(\br) \; +
$$
\be
\label{83}
+\; \Phi_0 \; \left\{ \left[ \; N_0 \; | \; \vp_n(\br) \; |^2 + 
2\rho_1^{(n)}(\br) \; \right] \; \vp_n(\br) + \sgm_1^{(n)}(\br) \; \vp_n^*(\br) 
+ \frac{\xi_1^{(n)}(\br)}{\sqrt{N_0}} \right\} \; ,
\ee
we obtain the eigenvalue equation
\be
\label{84}
H_{sup}[\; \vp_n\; ] \; \vp_n(\br) \; = \; E_n \; \vp_n(\br)
\ee
for the coherent modes $\varphi_n$. Here, the functions $\rho_1^{(n)}$, $\sigma_1^{(n)}$, 
and $\xi_1^{(n)}$ are the solutions to the equations where the role of the condensate 
function is played by the mode (\ref{81}).     

The condensate wave function can be represented as the expansion over the coherent 
modes:
\be
\label{85}
 \eta(\br,t) \; = \; 
\sqrt{N_0} \; \sum_n B_n(t) \; \vp_n(\br) \; e^{-i\om_n t} \; ,
\ee
where $B_n(t)$ is a slow function as compared to the fast oscillating exponential, and 
\be
\label{86}
\om_n \; \equiv \; E_n - \mu_0 \;   .
\ee
We assume that the external modulating field (\ref{4}) is in resonance with a chosen 
mode with the frequency, say $\omega_2$, so that the resonance condition (\ref{12}) is 
valid, in which now
$$
\Dlt \om \; = \; \om - \om_{21} \; , \qquad E_1 \; \equiv \; 
\min_n E_n \; = \; \mu_0 \; ,
$$
\be
\label{87}
\om_{21} \; = \; \om_2 - \om_1 \; = \; E_2 - E_1 \; = \; \om_2 \; .
\ee

We assume that at the initial moment of time the system is in its equilibrium state, 
so that only the ground-state coherent mode, with the energy level $E_1$ is occupied. 
Substituting expansion (\ref{85}) into equation (\ref{75}), we keep in mind the case of 
slowly varying in time $\rho_1$ and $\sigma_1$. Multiplying equation (\ref{75}) from the 
left by $\varphi_n^*(\br) \exp(i\omega_2) t$, averaging over time and integrating over
the spatial variable ${\bf r}$, we come to the equations
$$
i\; \frac{d B_1}{dt}\; = \; \al_{12} \; |\; B_2 \; |^2 \; B_1 +
\frac{1}{2} \; \bt_{12} \; B_2 \; e^{i\Dlt\om \cdot t} + \gm_1 \; B_1 \; ,
$$
\be
\label{88}
i\; \frac{d B_2}{dt}\; = \; \al_{21} \; |\; B_1 \; |^2 \; B_2 +
\frac{1}{2} \; \bt_{12}^* \; B_1 \; e^{-i\Dlt\om \cdot t} + \gm_2 \; B_2 \; ,
\ee
in which $B_n = B_n(t)$ and

\be
\label{89}
\gm_n \; = \; \Phi_0 \int \vp_n^*(\br) \; \left\{\; 2 [\; \rho_1(\br,t) -
 \rho_1^{(n)}(\br) \; ] \; \vp_n(\br) - \sgm_1^{(n)}(\br) \; \vp_n^*(\br) - \;
\frac{\xi_1^{(n)}(\br)}{\sqrt{N_0}} \; \right\} \; d\br \;   .
\ee
Recall that $\omega_1 = E_1 - \mu_0 = 0$, when $E_1$ corresponds to the lowest energy level.  

Employing the representation
\be
\label{90}
B_n \; = \; C_n \; e^{-i\gm_n t}
\ee
reduces equations (\ref{88}) to the form
$$
i\;\frac{dC_1}{dt} \; = \; \al_{12} \; |\; C_2 \;|^2 \; C_1 + 
\frac{1}{2} \; \bt_{12} \; C_2 \; e^{i\Dlt_{12} t} \; ,
$$
\be
\label{91}
i\;\frac{dC_2}{dt} \; = \; \al_{21} \; |\; C_1 \;|^2 \; C_2 + 
\frac{1}{2} \; \bt_{12}^* \; C_1 \; e^{-i\Dlt_{12} t} \;   ,
\ee
where
\be
\label{92}
 \Dlt_{12} \; \equiv \; \Dlt\om + \gm_1 - \gm_2 \;  .
\ee
 
In this way, we come to the equations (\ref{91}) that are similar to equations (\ref{16}),
although with the following difference. The energy levels $E_n$ in the eigenvalue equation 
(\ref{84}) are defined by an essentially more complicated expression (\ref{83}) and the 
detuning $\Delta_{12}$ in (\ref{91}) includes now the terms $\gamma_n$, depending on the 
interaction strength, as compared to the detuning $\Delta$ in (\ref{16}) independent of 
interactions. The larger detuning can destroy the resonance conditions, thus making 
impossible the resonant generation of the upper coherent modes. Nevertheless, since the 
terms $\gamma_1$ and $\gamma_2$ enter in the combination $\gamma_1 - \gamma_2$, they can
compensate each other making their difference much smaller than $\gamma_n$ itself. So, it 
looks to be not impossible to generate the coherent modes even in a system with rather 
strong interactions, provided the compensation effect is present.     

The treated case of relatively strong interactions reduces to that of weak interactions
if the interaction strength is small, such that the characteristics of non-condensed 
particles $\rho_1$, $\sigma_1$, and $\xi_1$ tend to zero. Then $\gamma_n$ also tends to zero.
Hence $C_n$ becomes $B_n$, and equation (\ref{88}) for $B_n$ reduces to equation (\ref{16})
for $c_n$.    

\section{Generalizations and Extensions}

The resonant generation of coherent modes, described above, is realized by means of the
trapping potential modulation with the frequency in resonance with the transition frequency
between two coherent modes. In this process, there occur several dynamical regimes and 
interesting dynamical transitions between these regimes. In addition, there may happen  
a number of other nontrivial dynamical effects some of which are surveyed below.

\subsection{Modulation through Interactions}

The other way of generating the coherent modes is by modulating particle interactions, which
can be accomplished by modulating an external magnetic field employed in Feshbach resonance
\cite{Timmermans_68}. An external magnetic field $B = B(t)$ influences the effective 
scattering length
\be
\label{93}
 a_s(B) \; = \; a_s \;\left( 1 - \; \frac{\Dlt B}{B-B_{res}} \right) \;  ,
\ee
in which $a_s$ is the scattering length far outside of the resonance field $B_{res}$ and 
$\Delta B$ is the resonance width. Then the interaction potential reads as
\be
\label{94}
 \Phi_s(t) \; = \; 4\pi \;\frac{a_s(B)}{m} \;  .
\ee
If the magnetic field oscillates around $B_0$ as 
\be
\label{95}
 B(t) \; = \; B_0 + b(t) \;  ,
\ee
with a small amplitude 
\be
\label{96}
 b(t) \; = \; b_1 \cos( \om t) + b_2 \sin(\om t) \;  ,
\ee
then the effective interaction takes the form
\be
\label{97}
 \Phi(t) \; \cong \; \Phi_0  + \Phi_1 \cos( \om t) + \Phi_2 \sin(\om t) \; ,
\ee
with
$$
\Phi_0 \; = \; \frac{4\pi}{m} \; a_s \;\left( 1 - \; 
\frac{\Dlt B}{B_0-B_{res}} \right) \;
$$
$$
\Phi_1 \; = \; \frac{4\pi a_s b_1 \Dlt B}{m(B_0-B_{res})^2} \; ,
\qquad
\Phi_2 \; = \; \frac{4\pi a_s b_2 \Dlt B}{m(B_0-B_{res})^2} \;  .
$$
The generation of coherent modes by the interaction modulation is similar to their generation
by the trap modulation \cite{Yukalov_12,Yukalov_14,Yukalov_50,Yukalov_69,Ramos_70}.

\subsection{Multi-Mode Generation}

In general, it is possible to generate not merely a single coherent mode but several modes
by applying several alternating fields, for instance, by applying the multi-frequency field
\be
\label{98}
V(\br,t) \; = \; \frac{1}{2} \sum_j \left[ \; B_j(\br) \; e^{i\om_j t} + 
B_j^*(\br) \; e^{-i\om_j t} \; \right] \; ,
\ee
in which the frequencies are in resonance with the required transition frequencies $\om_{mn}$. 
For example, two upper modes can be generated by using two alternating fields with the 
frequencies $\omega_1$ and $\omega_2$. Then, similarly to optical schemes \cite{Mandel_36},
three coherent modes can coexist, when different types of mode generation are used. In the 
cascade generation, one uses the resonance conditions
\be
\label{99}
 \om_1 \; = \; \om_{21} \; , \qquad \om_2 \; = \; \om_{32} \qquad 
(cascade ~ scheme) \;  ,
\ee
in the $V$-type scheme, the conditions
\be
\label{100}
\om_1 \; = \; \om_{21} \; , \qquad \om_2 \; = \; \om_{31} \qquad 
(V-type ~ scheme) \;   ,
\ee
and in the $\Lambda$-type scheme, the resonance conditions
\be
\label{101}
\om_1 \; = \; \om_{31} \; , \qquad \om_2 \; = \; \om_{32} \qquad 
(\Lbd-type ~ scheme) \; ,
\ee
are used. 

By varying the system parameters, various dynamic regimes can be realized exhibiting 
quasi-periodic oscillations \cite{Yukalov_14,Yukalov_73}. Contrary to the two-mode case, 
three or more coexisting modes can develop chaotic motion, when the strength of a 
generating field becomes sufficiently large, such that
\be
\label{102}
 \left| \; \frac{\bt_{mn}}{\al_{mn} } \; \right| \; \geq \; 0.639448 \; .
\ee

\subsection{Higher-Order Resonances}

Except the standard resonance, with $\om = \om_{21}$, there appear as well higher-order
resonances occurring under the effects of harmonic generation, when
\be
\label{103}
n \om \; = \; \om_{21} \qquad ( n = 1,2,\ldots )
\ee 
and under parametric conversion, when
\be
\label{104}
 \sum_j ( \pm \om_j) \; = \; \om_{21} \;  .
\ee
These effects require relatively strong generating field \cite{Yukalov_14,Yukalov_73}.

\section{Nonresonant Excitation}

The generation of coherent modes, considered above, requires the use of resonance, or 
quasi-resonance, conditions. Then there appear the following natural questions. First,
how long can the required resonance conditions be supported, not being spoiled by the 
effect of power broadening? Second, what is the influence of external noise on the 
dynamics of coherent modes? Third, can the coherent modes be generated without resorting 
to resonances, but merely applying a sufficiently strong external modulating field?

\subsection{Power Broadening}

The existence of the effect of power broadening does pose the limit to the ability of 
supporting the generation of coherent modes even in the case of pure resonance, since, 
in addition to resonant transitions there always occur non-resonant transitions, although 
their probability is small, however their effect accumulates with time. The interval of 
time, when, even under well defined resonance between two coherent modes, the generation 
of these modes can be realized, but after which the resonant generation becomes impossible, 
is found \cite{Yukalov_32} to be
\be
\label{105}
t_{mn} \;= \; \frac{\al_{mn}^2+\om_{mn}^2}{\bt_{mn}^2 \om_{mn} } \;   .
\ee
For typical traps, this time is of order $10-100$ s, which is quite long, being 
comparable to the typical lifetime of atoms in a trap \cite{Dalfovo_2,Courteille_3}. 
Recently \cite{Zhang_74} the lifetime of trapped atoms was shown to allow for an extension 
of up to $50$ minutes. 

The existence of external noise, of course, introduces irregularity in the dynamics of 
coherent modes, but if the noise is not too strong, it does not essentially disturbs the 
overall dynamical picture \cite{Yukalov_75}.    
  
Non-resonant alternating field can also generate coherent modes, provided the energy pumped 
into the system becomes sufficient for this mode generation. The energy per particle, 
injected into the system during the time period $t$ reads as
\be 
\label{106}
E_{inj} \; = \; \frac{1}{N} \int_0^t 
\left| \; \frac{\prt\lgl \; \hat H \; \rgl}{\prt t} \; \right| \; dt \; ,
\ee
where $\hat{H}$ is the energy Hamiltonian. A mode $n$ can be generated when the injected
energy surpasses the $n$-th mode energy, $E_{inj} \geq E_n$.

\subsection{Nonequilibrium-State Characteristics}

Different nonequilibrium states, comprising different modes, can be classified 
\cite{Yukalov_14,Yukalov_76} by the {\it effective temperature}
\be
\label{107}
 T_{eff} \; \equiv \; \frac{2}{3} \; [ \; E_{kin}(t) - E_{kin}(0) \; ] \;  ,   
\ee
expressed through the difference of kinetic energies at time $t$ and at the initial state, 
or by the {\it effective Fresnel number}
\be
\label{108}
T_{eff} \; \equiv \; \frac{\pi R^2}{\lbd_{eff} L}  \qquad
\left( \lbd_{eff} \equiv \sqrt{ \frac{2\pi}{m T_{eff}} } \right) \; ,
\ee
where $R$ and $L$ are the radius and length of the trap. 

Qualitatively different nonequilibrium states, displaying the appearance of different 
coherent modes, were studied in the experiments with trapped $^{87}$Rb atoms
\cite{Henn_77,Shiozaki_78,Seman_79} and in computer simulations \cite{Yukalov_80,Yukalov_81}, 
both being in good agreement with each other. The observed sequence of nonequilibrium 
states is listed in Table 1, where the numbers correspond to the lower threshold for the
appearance of the related states. Temperature is given in units of the transverse trap 
frequency $\omega_x=2\pi \times 210$ Hz employed in experiments 
\cite{Henn_77,Shiozaki_78,Seman_79} and in computer simulations \cite{Yukalov_80,Yukalov_81}.  

\begin{table}[ht]
\caption{Nonequilibrium states of a trapped Bose-Einstein condensate, characterized 
by the effective temperature $T_{eff}$ and effective Fresnel number $F_{eff}$. The
effective temperature is measured in units of the transverse trap frequency $\om_x$.}
\vskip 5mm
\centering
\renewcommand{\arraystretch}{1.2}
\begin{tabular}{|c|c|c|} \hline
                     &  $T_{eff}$  &  $F_{eff}$     \\ \hline
Weak nonequilibrium  &    0        &  0     \\ 
Vortex germs         &    0.29     &  0.11  \\ 
Vortex rings         &    1.21     &  0.23  \\ 
Vortex lines         &    2.26     &  0.31   \\
Vortex turbulence    &    5.54     &  0.49  \\ 
Droplet turbulence   &    8.56     &  0.61   \\
Wave turbulence      &    23.5     &  1.01    \\ \hline
\end{tabular}
\end{table}

By increasing the amount of energy injected into the trap, the system passes through several 
dynamical regimes with quite distinct properties. The sequence of the regimes is as follows.

(i) {\it Weak nonequilibrium}. At the beginning of the pumping procedure, there are no 
topological coherent modes, but there occur only elementary excitations describing density 
fluctuations. 

(ii) {\it Vortex germs}. Then, when the injected energy is not yet sufficient for the 
generation of the whole vortex rings, there arise vortex germs reminding broken pieces of 
vortex rings. 

(iii) {\it Vortex rings}. With the increasing injected energy, the whole rings appear in 
pairs, having the typical ring properties 
\cite{Iordanskii_82,Amit_83,Roberts_84,Jones_85,Jackson_86,Barenghi_87,Reichl_88}. 

(iv) {\it Vortex lines}. At the next stage, there appear the pairs of vortex lines 
\cite{Pethick_89}. They arise in pairs, since no rotation is imposed on the system, so that 
the total vorticity has to be zero. 

(v) {\it Vortex turbulence}. Upon generating a large number of vortices, there develops 
the regime of quantum vortex turbulence. Because of the absence of any imposed anisotropy, 
the vortices form a random tangle characteristic of the Vinen turbulence 
\cite{Vinen_90,Vinen_91,Vinen_92,Tsubota_93,Nemirovskii_94,Tsatos_95}, 
as opposed to the Kolmogorov turbulence of correlated vortex lines \cite{Nemirovskii_94}.
Several specific features confirm the existence of quantum vortex turbulence. Thus, when 
released from the trap, the atomic cloud expands isotropically, which is typical of
Vinen turbulence \cite{Henn_77,Shiozaki_78,Seman_79}. The radial momentum distribution, 
obtained by averaging in the axial direction, exhibits a specific power law typical of an 
isotropic turbulent cascade \cite{Zakharov_96,Thompson_97,Navon_98}. The system relaxation 
from the vortex turbulent state displays a characteristic universal scaling \cite{Moreno_99}. 

(vi) {\it Droplet turbulence}. Increasing further the amount of the injected energy by a
longer pumping or by rising the amplitude of the alternating field transforms the system
into an ensemble of coherent droplets floating in a see of an uncondensed cloud. The density 
of the coherent droplets is around $100$ times larger than that of their incoherent 
surrounding. Each droplet consists of about $40$ atoms. The lifetime of a droplet is of the
order of $0.01$ s. This state can be called droplet turbulence, or grain turbulence 
\cite{Yukalov_76,Yukalov_80,Yukalov_81}. 

(vii) {\it Wave turbulence}. When coherence in the system is completely destroyed, the 
system enters the regime of wave turbulence which is the regime of weakly nonlinear 
dispersive waves \cite{Nazarenko_100}. Strictly speaking, the transformation of the droplet 
turbulence into wave turbulence is not a sharp transition but a gradual crossover. The 
transition point is conditionally accepted as the point where the number of coherent droplets 
diminishes by half.     

When a nonequilibrium system relaxes to its equilibrium state, passing from a state with
a symmetry to the state where the symmetry becomes broken, it passes though the stage with
the appearing topological  defects, such as grains, cells, vortices, strings, and like that. 
This is called the Kibble-Zurek mechanism \cite{Kibble_101,Kibble_102,Zurek_103,Zurek_104}.  
In our experiments and computer modeling \cite{Henn_77,Shiozaki_78,Seman_79,Yukalov_80,
Yukalov_81}, we follow the opposite way by transforming an equilibrium system with broken 
global gauge symmetry to a nonequilibrium gauge-symmetric system, passing through the stages 
of arising topological defects that are vortex germs, vortex rings, vortex lines, and coherent 
droplets. Therefore, this opposite way can be named the {\it inverse Kibble-Zurek scenario} 
\cite{Yukalov_81}.

\subsection{Dynamic Scaling}

Nonequilibrium regimes can be distinguished and characterized by scaling laws. It is known 
that many dynamic systems exhibit a kind of self-similarity in their evolution. This was 
noticed by Family and Vicsek \cite{Viscek_105,Family_106} in the process of diffusion-limited 
aggregation of clusters in two dimensions. The Family-Vicsek dynamic scaling, describes 
the behavior of a probability distribution $f(x,t)$ of a variable $x$ at different instants 
of time $t$, so that
\be
\label{109}
 f(x,t) \; = \; \left( \frac{t}{t_0} \right)^\al
F\left( x\left( \frac{t}{t_0}\right)^\bt , t_0 \right) \; 
\left( \frac{x}{x_0} \right)^\gm \;  ,
\ee
where $F(x,t)$ is a universal function, $x_0$ and $t_0$ are fixed reference values, and 
$\alpha$, $\beta$, and $\gamma$ are universal scaling exponents. Numerous cases of 
nonequilibrium dynamics display scaling laws, for instance, polymer degradation \cite{Ziff_107}, 
kinetics of aggregation \cite{Dongen_108,Kreer_109,Hassan_110,Hassan_111,Hassan_112}, complex 
networks \cite{Hassan_113}, growth models \cite{Kardar_114,Souza_115}, fractional Poisson 
processes \cite{Kreer_116}, and other dynamical processes \cite{Leonel_117}. Scaling laws and
universal critical exponents appear in the theory of nonthermal fixed points when the system 
is far from equilibrium \cite{Pineiro_118,Schmied_119,Mikheev_120}, which distinguishes the
scaling-law regime from the quasi-stationary stage of prethermalization 
\cite{Berges_121,Gring_122,Langen_123,Mori_124}. Cold trapped Bose gas serves as a very
convenient object for studying quantum turbulence \cite{Henn_77,Shiozaki_78,Seman_79,
Yukalov_80,Yukalov_81,Tsubota_93,Nemirovskii_94,Tsatos_95,Thompson_97,Navon_98,Moreno_99}
and its relaxation \cite{Moreno_99,Nowak_125,Kwon_126,Nowak_127,Madeira_128}. 

Distinct stages in the relaxation dynamics of a harmonically trapped three-dimensional
Bose-Einstein condensate of $^{87}$Rb, driven to a turbulent state by an external oscillating 
field, is analyzed in the paper \cite{Moreno_99}. The angular-averaged two-dimensional 
momentum distribution $n(k,t)$ is measured, for small momenta $k \ra 0$, in the time-of-flight 
experiment \cite{Garcia_129}. The universal dynamical scaling in the time evolution of the 
momentum distribution is observed:
\be
\label{110}
n(k,t) \; = \; \left( \frac{t}{t_0} \right)^\al
n \left( k \left( \frac{t}{t_0}\right)^\bt , t_0 \right) \; ,
\ee
where $t_0$ is an arbitrary reference time within the temporal window where the scaling is
observed. The universal exponents are: 
$$
 \al \;= \; - 0.5 \; , \qquad \bt \; = \; - 0.25 \;  .
$$  
This universal scaling (\ref{110}) corresponds to a {\it direct energy cascade} from the 
low-momentum to the high-momentum states, when the condensate becomes depleted. 

Then, after a {\it prethermalization stage}, there appears an {\it inverse energy cascade} 
from the high-momentum to the low-momentum states, which implies the repopulation of the 
condensate. At the condensate revival stage, the dynamic scaling has the form
\be
\label{111}
n(k,t) \; = \; \left( \frac{t_b-t}{t_b-t_0} \right)^\lbd
n \left( k \left( \frac{t_b-t}{t_b-t_0}\right)^\mu , t_0 \right) \;   ,
\ee
with the universal exponents
$$
 \lbd \;= \; - 1.5 \; , \qquad \mu \; = \; - 0.9 \;  ,
$$
showing that the condensate fraction sharply increases, which is called \cite{Zhu_130,Zhu_131}
the condensate blowup.

\section{Conclusion}

Dynamic transitions between different nonequilibrium states of trapped Bose-Einstein 
condensates, subject to the action of alternating fields, are surveyed. The applied external 
fields can be of two types, resonant and nonresonant. A resonant field implies that its
frequency is tuned close to a resonance with some transition frequency of the trapped system.
The transition frequency is the difference between two chosen energy levels of the trapped 
system. Several external alternating fields, with different frequencies can also be used.
Resonant fields do not need to be strong. More important is the presence of resonance 
conditions. 

Resonant fields generate nonlinear coherent modes in trapped condensates. Depending on the
ratio between the amplitude of the alternating field and the interaction strength of atoms,
there can appear several dynamic states, including the mode-locked Josephson regime, critical
dynamics, mode-unlocked Josephson regime, pitchfork bifurcation, and Rabi regime. The dynamic
transition, occurring on the critical line in the effect of separatrix crossing reminds a 
phase transition in equilibrium statistical systems. These dynamical transitions can be realized 
in weakly interacting trapped Bose gases. We show that the generation of coherent modes and 
the related dynamic transitions can, in principle, be implemented in strongly interacting 
superfluids too, although it is a much more complicated task. 

Employing several alternating fields, it is possible to generate several coherent modes and
realize higher-order resonance phenomena, such as harmoic generation and parametric 
conversion.

The other way of generating nonequilibrium states in trapped Bose condensates is through the
use of sufficiently strong nonresonant fields. Then one can produce a sequence of 
nonequilibrium states containing vortex germs, vortex rings, and vortex lines, as well as 
generate different turbulent regimes, such as vortex turbulence, droplet turbulence, and wave
turbulence. Nonequilibrium states of superfluids can be characterized by effective temperature
and effective Fresnel number. Different stages of nonequilibrium systems can be distinguished 
by the existence of specific dynamic scaling.

\vskip 3mm

{\bf Declarations}

\vskip 2mm

{\bf Author contributions}: All authors contributed to the study conception and design. 
Material preparation, data collection and analysis were performed by V.I. Yukalov and 
E.P. Yukalova. The first draft of the manuscript was written by V.I. Yukalov and all 
authors commented on previous versions of the manuscript. All authors read and approved 
the final manuscript.

\vskip 2mm

{\bf Funding}: No funding was received to assist with the preparation of this manuscript.

\vskip 2mm

{\bf Financial interests}: The authors have no relevant financial or non-financial 
interests to disclose. 

\vskip 2mm

{\bf Competing interests}: The authors have no competing interests to declare that are 
relevant to the content of this article.

\end{document}